\documentclass[a4paper,fleqn,usenatbib]{mnras}

\usepackage{newtxtext,newtxmath}

\usepackage[T1]{fontenc}
\usepackage{ae,aecompl}

\usepackage{graphicx}	
\usepackage{amsmath}	

\newcommand{\vsini}{$v_{\rm e} \sin i$}
\newcommand{\kms}{km\,s$^{-1}$}
\newcommand{\ms}{m\,s$^{-1}$}

\newcommand{\bz}{$\langle B_{\rm z}\rangle$}
\newcommand{\dra}{$\varphi$~Dra}

\newcommand{\figps}[3]{\resizebox{#1}{!}{\rotatebox{#2}{\includegraphics{#3}}}}

\title[Surface structure and variability of $\varphi$~Dra]{Magnetic field topology, chemical spot distributions and photometric variability of the Ap star $\varphi$~Draconis}

\author[O. Kochukhov et al.]
{O.\ Kochukhov$^1$\thanks{E-mail: oleg.kochukhov@physics.uu.se},
N.\ Papakonstantinou$^1$,
C.\ Neiner$^2$,
\\
$^1$Department of Physics and Astronomy, Uppsala University, Box 516, Uppsala 75120, Sweden \\
$^2$LESIA, Paris Observatory, PSL University, CNRS, Sorbonne Universit\'e, Universit\'e de Paris, 5 place Jules Janssen, 92195 Meudon, France
}

\date{Accepted 2022 January 6. Received 2022 January 5; in original form 2021 December 17}

\pubyear{2022}

\begin{document}
\label{firstpage}
\pagerange{\pageref{firstpage}--\pageref{lastpage}}
\maketitle

\begin{abstract}
The primary component of the multiple star $\varphi$~Dra is one of the brightest magnetic chemically peculiar stars in the northern sky. Here we report results of a comprehensive study of the rotational photometric variability, binarity, magnetic field geometry, and surface chemical spot structure for this star. We derived a precise photometric rotational period of 1.71650213(21)~d based on one year of \textit{TESS} nearly continuous space observations and discovered modulation of the stellar light curve with the phase of the 127.9-d binary orbit due to the light time travel effect. We revised parameters of the binary orbit and detected spectroscopic contribution of the secondary. A tomographic mapping technique was applied to the average intensity and circular polarisation profiles derived from Narval high-resolution spectropolarimetric observations. This analysis yielded a detailed map of the global magnetic field topology together with the surface distributions of Si, Cr, and Fe abundances. Magnetic mapping demonstrates that the surface field structure of $\varphi$~Dra is dominated by a distorted dipolar component with a peak field strength of 1.4~kG and a large asymmetry between the poles. Chemical maps show an enhancement of Cr, Fe and, to a lesser extent, Si in a series of spots encircling intersections of the magnetic and rotational equators. These chemical spot geometries do not directly correlate with either the local field strength or the field inclination.
\end{abstract}




\section{Introduction}
\label{sec:intro}

The upper main sequence magnetic chemically peculiar (ApBp or mCP) stars are A- and B-type stars distinguished by large under- and overabundances of chemical elements in their surface layers and the presence of a strong, stable, globally organised magnetic field. This chemical and magnetic structure is frozen into the stellar plasma and rotates with the star, giving rise to a characteristic synchronous rotational variability of the brightness in most photometric passbands, spectral energy distribution, individual line profiles, and magnetic observables. This unique combination of unusual properties and pronounced variability makes ApBp stars very useful, and sometimes the only suitable objects, for detailed studies of the origin, structure and evolution of the fossil magnetic fields in radiative stellar envelopes (e.g. \citealt{braithwaite:2006,landstreet:2007}; \citealt*{kochukhov:2019}), an interplay between the mass loss and rotational spin down (e.g. \citealt*{ud-doula:2009}; \citealt{shultz:2019a}), and microscopic chemical transport processes responsible for the observed chemical anomalies and starspot formation (e.g. \citealt*{michaud:2015}; \citealt{kochukhov:2017,ryabchikova:2017}).

With a $V$ magnitude of 4.22, \dra\ (43~Dra, HD\,170000, HR\,6920, HIP\,89908, TIC\,356822358) is one of the brightest magnetic chemically peculiar stars. It is a rapidly rotating object showing a coherent spectroscopic \citep{jamar:1977,kuschnig:1998}, photometric (\citealt{musielok:1980,prvak:2015}; \citealt*{bernhard:2020}), and magnetic \citep{landstreet:1977,sikora:2019a} variability with a period of $\approx$\,1.72~d. This star is, in fact, a triple system with the visual secondary (component B) orbiting the primary (component A) on a 308-yr orbit with a semi-major axis of $<$\,$1\arcsec$ \citep{liska:2016}. The magnitude difference between these two components is measured to be 1.3--1.5~mag depending on wavelength \citep{fabricius:2000,rutkowski:2005}. The primary is itself a single-line eccentric spectroscopic binary (components Aa and Ab) with a period of 128~d \citep{liska:2016,bischoff:2017}. Multiplicity of \dra\ is commonly ignored by studies of the magnetic Ap star (component Aa) since no evidence of the spectroscopic contribution of components Ab or B has been reported in the literature.

The global magnetic field of \dra\ was first detected by \citet{landstreet:1977} using the Balmer-line mean longitudinal magnetic field, \bz, measurements with a typical uncertainty of 100~G. That study established a smooth, sinusoidal magnetic field curve with \bz\ changing from about $-300$ to $+500$~G and interpreted these observations with a 3~kG centred dipolar magnetic field topology with a low obliquity ($\beta\le20\degr$) relative to the rotational axis. Considerably more precise longitudinal magnetic field measurements were obtained for \dra\ by \citet{sikora:2019a} from metal lines in high-resolution circular polarisation spectra. Although the inferred longitudinal field curve was fully compatible with the earlier hydrogen line results, \citet{sikora:2019a} derived significantly different values of the dipolar field strength and obliquity -- 1.75~kG and 70\degr\ -- compared to the dipolar model proposed by \citet{landstreet:1977}.

Surface distributions of several chemical elements on the surface of \dra\ were studied by \citet{kuschnig:1998} with the help of the Doppler imaging (DI) technique. Using these chemical spot maps, \citet{prvak:2015} calculated theoretically expected flux variations in different photometric bands and succeeded in reproducing the observed UV and optical photometric behaviour of this star. \dra\ was included in the volume-limited survey of magnetic Ap stars by \citet{sikora:2019}. These authors derived atmospheric parameters, assessed evolutionary state, and measured mean abundances. They have concluded that \dra\ is a 3.56 $M_\odot$ star that has completed 88 per cent of its life on the main sequence.

Motivated by availability of new high-quality spectropolarimetric observations and unique long-duration, precise space-borne photometric data for \dra, we have carried out a comprehensive study of the magnetic field topology, chemical abundance spots, photometric variability, and binarity of this star. Our paper is organised as follows. Sect.~\ref{sec:obs} presents observational data, followed by the description of analysis and results in Sect.~\ref{sec:res}, including a revision of the orbit of the inner spectroscopic binary (Sect.~\ref{sec:binary}), time series analysis of the space photometric data (Sect.~\ref{sec:tess}), assessment of the mean polarisation profiles (Sect.~\ref{sec:lsd}), and Zeeman Doppler imaging (Sect.~\ref{sec:zdi}). The paper ends with conclusions and discussion in Sect.~\ref{sec:disc}.

\section{Observational data}
\label{sec:obs}

\subsection{Spectropolarimetry}

In this study we analysed 22 high-resolution circular polarisation observations of \dra\ collected with the Narval spectropolarimeter installed at the 2m T\'elescope Bernard Lyot (TBL) at the Pic du Midi observatory in France. The same data set was previously studied by \citet{sikora:2019,sikora:2019a}. These observations provide a continuous coverage of the 3600--10\,000~\AA\ wavelength interval at a resolving power of $R\approx65\,000$. The spectra of \dra\ were obtained between August 2016 and February 2017. Each observation consisted of four 360~s sub-exposures acquired with different polarimeter configurations, allowing one to derive Stokes $V$ and diagnostic null spectra \citep{donati:1997,bagnulo:2009}. The optimal extraction of one-dimensional spectra and their polarimetric demodulation was carried out with an updated version of the {\sc esprit} reduction code \citep{donati:1997}, yielding data with a typical signal-to-noise ratio ($S/N$) of 900--1200 per 1.8~\kms\ velocity bin at $\lambda=5500$~\AA. The spectra were post-processed with the continuum normalisation routines described by \citet{rosen:2018}. Further details on individual observations, including observing dates, heliocentric Julian dates, and $S/N$ values are given in the first three columns of Table~\ref{tab:obs}. 

\begin{table*}
\caption{Log of Narval spectropolarimetric observations of \dra. The columns give the UT and heliocentric Julian dates of mid-exposure for each observation, the $S/N$ per 1.8~\kms\ velocity bin at 5500~\AA, the rotational and orbital phases, the mean longitudinal magnetic field determined from LSD profiles, and the radial velocity measured from the hydrogen line cores. The average uncertainty of the latter measurements is 0.73~\ms. \label{tab:obs}}
\begin{tabular}{llrccrr}
\hline
UT date & HJD & $S/N$ & Rotational & Orbital & \multicolumn{1}{c}{\bz} & \multicolumn{1}{c}{$V_{\rm H}$} \\
        &     &       & phase      & phase   &  \multicolumn{1}{c}{(G)} & \multicolumn{1}{c}{(\kms)} \\
\hline
2016-08-20 & 2457621.401 &  986 & 0.348 & 0.293 & $-236\pm26$ & $-24.2$ \\
2016-08-21 & 2457622.368 & 1176 & 0.912 & 0.301 & $ 387\pm24$ & $-25.5$ \\
2016-08-22 & 2457623.417 &  891 & 0.523 & 0.324 & $-170\pm33$ & $-23.6$ \\
2016-08-23 & 2457624.421 & 1069 & 0.108 & 0.332 & $ 407\pm25$ & $-24.1$ \\
2016-08-24 & 2457625.414 & 1166 & 0.686 & 0.340 & $ 191\pm27$ & $-23.7$ \\
2016-08-29 & 2457630.448 &  970 & 0.619 & 0.074 & $  -2\pm32$ & $-21.0$ \\
2016-08-31 & 2457631.517 & 1068 & 0.242 & 0.363 & $-102\pm24$ & $-19.6$ \\
2016-09-02 & 2457634.387 & 1094 & 0.914 & 0.081 & $ 430\pm26$ & $-19.7$ \\
2016-09-03 & 2457635.367 &  947 & 0.485 & 0.371 & $ -76\pm30$ & $-17.7$ \\
2016-09-06 & 2457638.397 & 1033 & 0.250 & 0.089 & $ -82\pm24$ & $-16.9$ \\
2016-09-07 & 2457639.404 & 1073 & 0.836 & 0.097 & $ 308\pm28$ & $-17.8$ \\
2016-09-08 & 2457640.391 & 1142 & 0.411 & 0.601 & $-286\pm24$ & $-15.4$ \\
2016-09-11 & 2457643.392 & 1115 & 0.160 & 0.609 & $ 232\pm24$ & $-15.9$ \\
2016-09-12 & 2457644.310 & 1048 & 0.695 & 0.617 & $ 141\pm30$ & $-16.5$ \\
2016-12-11 & 2457734.237 &  834 & 0.084 & 0.624 & $ 437\pm33$ & $-43.2$ \\
2016-12-12 & 2457735.235 & 1057 & 0.666 & 0.192 & $ 109\pm30$ & $-40.7$ \\
2016-12-13 & 2457736.234 & 1106 & 0.248 & 0.199 & $-112\pm23$ & $-38.7$ \\
2016-12-14 & 2457737.236 & 1162 & 0.831 & 0.207 & $ 364\pm26$ & $-37.6$ \\
2017-02-17 & 2457801.722 & 1027 & 0.400 & 0.215 & $-291\pm26$ & $ -7.2$ \\
2017-02-18 & 2457802.710 & 1136 & 0.976 & 0.223 & $ 421\pm25$ & $ -9.2$ \\
2017-02-19 & 2457803.709 & 1075 & 0.557 & 0.262 & $ -68\pm28$ & $ -8.6$ \\
2017-02-20 & 2457804.699 &  944 & 0.134 & 0.271 & $ 350\pm28$ & $ -8.2$ \\
\hline
\end{tabular}
\end{table*}

\subsection{Space photometry}

The Transiting Exoplanet Survey Satellite \citep[\textit{TESS},][]{ricker:2015} was launched in April 2018 and is currently carrying out a survey of most of the sky in sectors measuring $24\degr\times96\degr$. Each sector is observed for 27.4~d, which corresponds to two orbits of the satellite. These observations yield continuous high-precision photometric data in a wide red (600--1000~nm) bandpass over $\approx$\,13~d time span corresponding to each orbit with short gaps between the orbits and sectors. The sectors are oriented along ecliptic longitudes, resulting in repeated observations of targets in the continuous viewing zones (CVZ) around both ecliptic poles. \dra\ is one of the few bright Ap stars located within the northern \textit{TESS} CVZ. This star was therefore observed with a 2-min cadence over the entire second year of the \textit{TESS} mission. Specifically, the data are available from sector 14 (June 2019) through sector 26 (June 2020), with the exception of sector 22 when the target was located at the edge of the \textit{TESS} field of view and a light curve could not be extracted. In total, 200456 photometric measurements with a typical precision of 0.065~mmag are available. These data covers a period of 351.8~d, which is equivalent to 205 stellar rotation periods and 2.75 orbital periods of the inner binary in the \dra\ system.

We downloaded reduced \textit{TESS} observations of \dra\ from the Mikulski Archive for Space Telescopes (MAST)\footnote{\url{https://mast.stsci.edu}}. Two versions of the light curves produced by the  Science
Processing Operations Center (SPOC) pipeline \citep{jenkins:2016} are available: the Simple Aperture Photometry (SAP) and the Pre-search Data Conditioning Simple Aperture Photometry (PDCSAP) data. The latter includes additional corrections intended to remove long-term trends and systematics from stellar light curves. However, similar to some other recent studies which analysed \textit{TESS} observations of bright Ap stars \citep{holdsworth:2021}, we found significant artefacts in the PDCSAP version of the light curve of \dra, rendering its time series analysis problematic. For this reason we chose to use the SAP light curves in this paper.

The \textit{TESS} cameras use a relatively coarse pixel scale of 21\arcsec\ and do not resolve the three components comprising the \dra\ system. A large numerical aperture of $7\times23$ pixel was adopted to extract the light curve for this bright star, resulting in signal integration over an area of $2.5\arcmin\times8\arcmin$. Concerned with possible contamination by additional nearby stars, we have examined {\it Gaia} eDR3 \citep{gaia-collaboration:2021} sources contributing to this aperture. No significant contaminants were found.

\section{Analysis and results}
\label{sec:res}

\subsection{Spectroscopic binary orbit}
\label{sec:binary}

The orbital radial velocity variation of \dra\ has been known for a long time \citep[e.g.][]{beardsley:1969,abt:1973}. However, the period and amplitude of the radial velocity curve were poorly constrained \citep{liska:2016} due to difficulty of extracting accurate velocity measurements from broad, variable spectral lines of the Ap star. This problem has been largely overcome by \citet{bischoff:2017}, who provided 62 precise radial velocity measurements obtained from hydrogen Balmer lines. Since stellar atmospheres are predominantly made of hydrogen, the lines of this element are much less affected by surface chemical abundance inhomogeneities compared to metal lines (typically Mg, Fe, Cr, Si) accessible for radial velocity measurements. Here we have replicated the methodology used by \citet{bischoff:2017} by calculating the centre-of-gravity velocities from the cores of H$\beta$, H$\gamma$, and H$\delta$ (H$\alpha$ was avoided due to contamination by variable telluric lines). The resulting average H-line radial velocities are reported in Table~\ref{tab:obs}. The scatter between different Balmer lines amounts to 0.6--0.8~\kms, which is much smaller compared to 5--10~\kms\ rotational jitter exhibited by metal lines.

\begin{figure}
\centering
\figps{\hsize}{0}{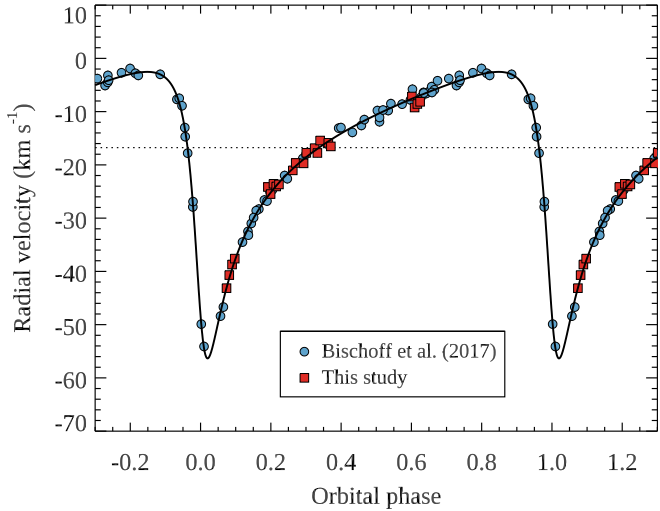}
\caption{Orbital radial velocity variation of \dra\ measured using hydrogen lines. Symbols show the radial velocity measurements obtained in this study (squares) and by \citet[][circles]{bischoff:2017}. The solid line shows the orbital fit discussed in the text.}
\label{fig:orbit}
\end{figure}

The 22 new radial velocity measurements were combined with the data from \citet{bischoff:2017} and the full set of SB1 spectroscopic orbital elements ($P_{\rm orb}$, HJD$_0$, $K$, $\gamma$, $e$, $\omega$) was obtained by fitting observations with a non-linear least-squares optimisation algorithm \citep{markwardt:2009}. The results are presented in Table~\ref{tab:orbit}. The observed radial velocities are compared to the radial velocity curve corresponding to the best-fitting orbital solution in Fig.~\ref{fig:orbit}. The standard deviation of our H-line radial velocity measurements around the fitted curve is 0.9~\kms, which is consistent with the uncertainty estimate above.

\begin{table}
\caption{Orbital parameters of the inner spectroscopic binary obtained from the radial velocity measurements of hydrogen lines and with the DI analysis of the Fe LSD profiles. \label{tab:orbit}}
\begin{tabular}{ll}
\hline
Parameter & Value \\
\hline
$P_{\rm orb}$  (d) &   $127.914\pm0.062$ \\
HJD$_0$ (d) & $2456957.31\pm0.18$ \\
$K_{\rm H}$ (\kms) &     $26.90\pm0.29$ \\
$\gamma_{\rm H}$ (\kms) &    $-16.76\pm0.11$ \\
$e$  &     $0.6725\pm0.0044$ \\
$\omega$ (\degr) &    $134.47\pm0.82$ \\
\hline
$K_{\rm Fe}$ (\kms) &     $33.87\substack{+0.25 \\ -0.17}$ \\
$\gamma_{\rm Fe}$ (\kms) &    $-16.44\substack{+0.10 \\ -0.08}$ \\
\hline
\end{tabular}
\end{table}

An important caveat of this orbital analysis is that it relies on the assumption that the H-line radial velocity variation is due to the orbital motion of the component Aa and that the secondary component of the inner binary, Ab, does not contribute to the spectra. But, as will be demonstrated below (Sect.~\ref{sec:lsd}), \dra\ is actually an SB2 system with the component Ab contributing to the metal line spectra and therefore to hydrogen lines as well. This means that the H-line radial velocity measurements yield a systematically underestimated orbital semi-amplitude. A corrected value of this parameter is derived in Sect.~\ref{sec:zdi} as part of the Doppler imaging analysis.

\subsection{Photometric variability}
\label{sec:tess}

\begin{figure}
\centering
\figps{\hsize}{0}{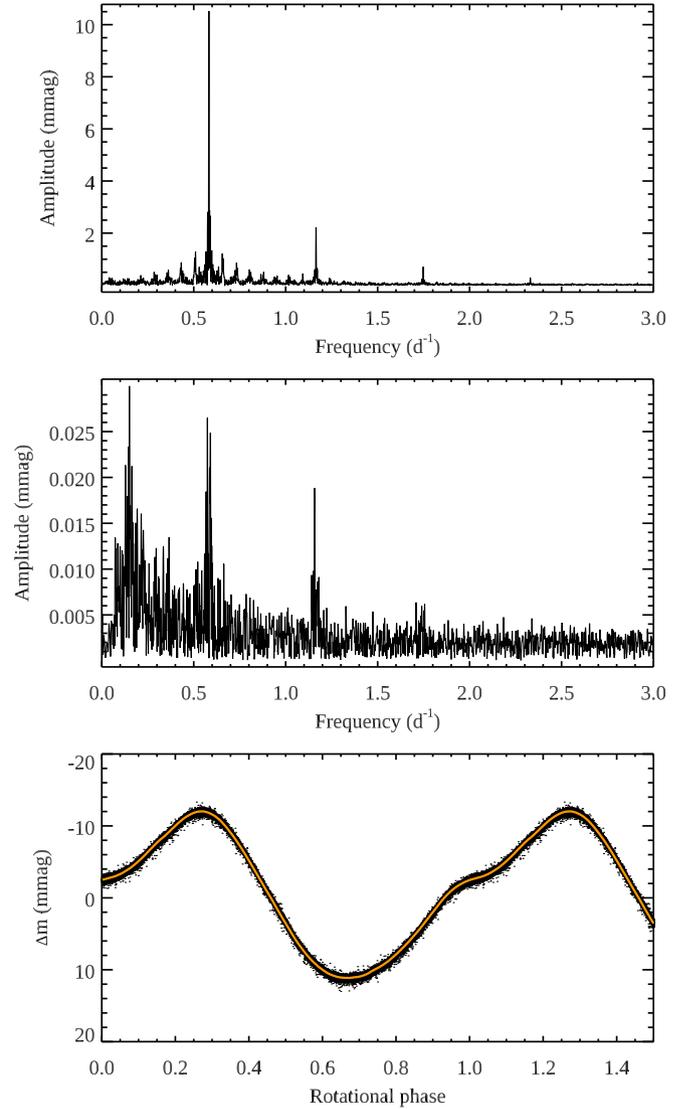}
\caption{Upper panel: periodogram for the \textit{TESS} light curve of \dra. Middle panel: periodogram for the residual time series after removing the rotational signal. Lower panel: \textit{TESS} observations plotted as a function of the rotational phase (symbols) along with the harmonic fit (solid line).}
\label{fig:tess}
\end{figure}

Time series analysis of the \textit{TESS} observations of \dra\ was carried out using standard methods which combined calculation of periodograms and fitting data in the time domain \citep[e.g.][]{kochukhov:2021b}. The generalised Lomb-Scargle periodogram \citep{zechmeister:2009} for the entire \textit{TESS} data set is presented in Fig.~\ref{fig:tess} (upper panel). The rotational frequency and its multiple harmonics are clearly visible. 

In order to extract the full information about rotational modulation we fitted the \textit{TESS} light curve with a superposition of 10 harmonics of the rotational frequency. This periodic model was added to third-degree polynomial functions describing slow instrumental trend within each of the 24 continuous 10.1--13.5~d light curve segments. All parameters of this composite model were optimised simultaneously with the help of a non-linear least-squares fitting algorithm. This analysis yields a stellar rotational period of 1.71650213(21)~d, which agrees with all previous determinations within uncertainties (see Table~\ref{tab:period}) but is considerably more precise owing to a high quality and long baseline of the \textit{TESS} photometric observations of \dra. Keeping the same reference Julian date as derived by \citet{sikora:2019a}, we adopted the following rotational ephemeris in this work
$$
HJD = 2442632.30626 + 1.71650213(21) \times E.
$$
The \textit{TESS} data phased according to this ephemeris is compared to the model light curve in Fig.~\ref{fig:tess} (bottom panel). The photometric variation of \dra\ in the \textit{TESS} passband is dominated by a sinusoidal (single-wave) component, but the contribution of higher order harmonics is also significant.

\begin{table}
\caption{Photometric rotational period determinations for \dra. \label{tab:period}}
\begin{tabular}{ll}
\hline
Period (d) & Reference \\
\hline
1.71646(6) &    \citet{musielok:1980} \\
1.716500(2)  &  \citet{prvak:2015} \\
1.71665(9) &    \citet{bernhard:2020} \\
1.7166(1)  &    \citet{paunzen:2021b} \\
1.71650213(21) & This work \\
\hline
\end{tabular}
\end{table}

The periodogram of the residual light curve obtained by subtracting the model of rotational variability from the observed data is shown in Fig.~\ref{fig:tess} (middle panel). There are clear residual signals associated with the rotational frequency and its harmonics. In particular, there are two frequency peaks separated by $\pm$0.0080(15) d$^{-1}$ from the rotational frequency. This frequency splitting is consistent with the orbital frequency 0.007818(4) d$^{-1}$ derived in Sect.~\ref{fig:orbit}. This suggests that the photometric rotational variability of \dra\ is modulated by the orbital motion within the Aab binary system. In addition, Fig.~\ref{fig:tess} reveals the presence of a signal with a period of 6.649(80)~d, which likely corresponds to the rotation of either component Ab or B.

\begin{figure}
\centering
\figps{\hsize}{0}{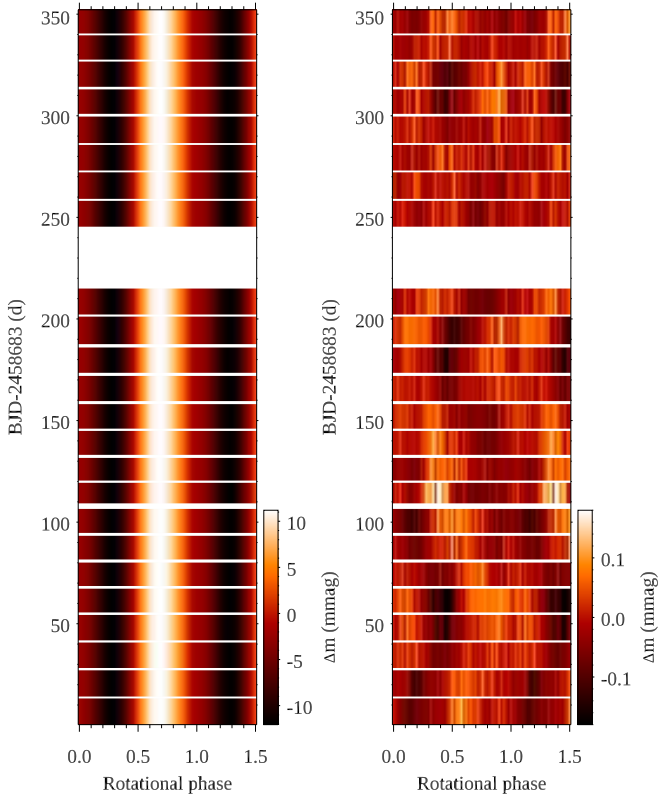}
\caption{Left panel: phased \textit{TESS} time series of \dra\ averaged over 10--13~d light curve segments. Right panel: residual light curves for each segment.}
\label{fig:stacked}
\end{figure}

To gain further insight into long-term changes of the photometric rotational modulation of \dra\ we constructed average binned light curves for each of the 24 continuous data segments. The time evolution of these average light curves is shown in Fig.~\ref{fig:stacked} (left panel). No significant changes on the time span of one year covered by \textit{TESS} photometry are immediately obvious. However, when we subtract the median light curve and examine evolution of the residuals (Fig.~\ref{fig:stacked}, right panel), a clear repetitive S-shaped pattern emerges. This cyclic structure in the residuals is particularly clear for the first 200 days of \textit{TESS} observations and appears to be consistent with $P_{\rm orb}\approx130$~d derived earlier.

\begin{figure}
\centering
\figps{\hsize}{0}{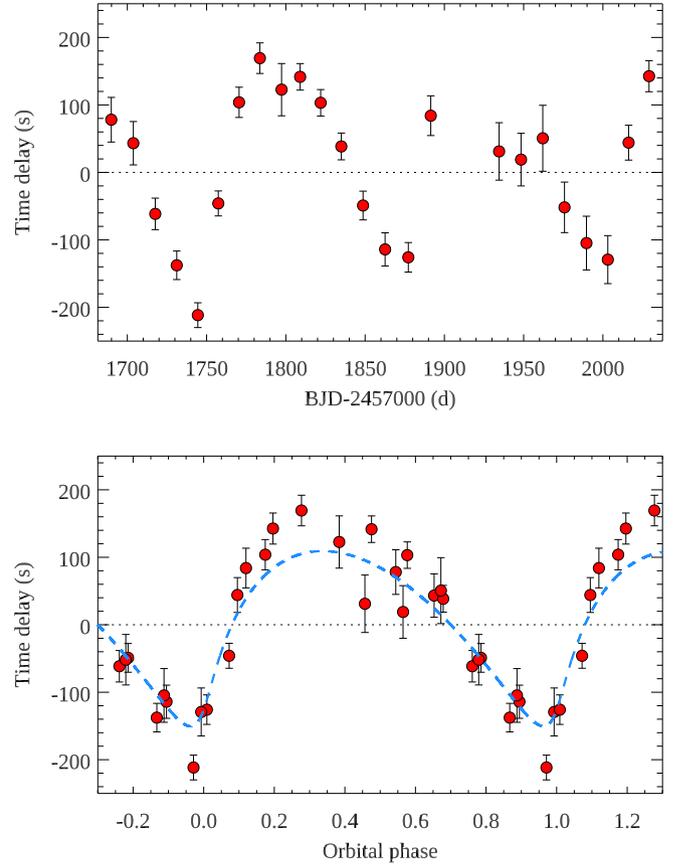}
\caption{Upper panel: time delay as a function of time for 10--13~d \textit{TESS} light curve segments. Lower panel: same data plotted as a function of the orbital phase. The dashed line in the lower panel shows the time delay predicted by the Fe-line spectroscopic orbital solution.}
\label{fig:delay}
\end{figure}

What phenomenon can be responsible for the apparent modulation of the rotational light curve on the orbital time scale? The light travel time effect \citep{shibahashi:2012,murphy:2014} seems to be the most plausible explanation, which requires no new physical processes or effects besides the orbital motion. Previous observations of pulsating stars in binary systems showed that the light travel time leads to time delays or phase shifts in pulsational light curves. The same principle can be applied to the rotational light curves of binary stars provided that $P_{\rm rot} \ll R_{\rm orb}$, which is satisfied for \dra. Following this line of reasoning, we expressed the orbital variation of \dra\ light curve in terms of time-delay measurements. This was accomplished by finding a time offset that would minimise discrepancy between the model light curve shown in Fig.~\ref{fig:tess} and observations in each of the 24 light curve segments. The resulting time delays amount to about $\pm150$~s and vary systematically with time (Fig.~\ref{fig:delay}, upper panel). A clear periodic structure becomes evident when the time-delay measurements are plotted as a function of the orbital phase (Fig.~\ref{fig:delay}, lower panel). 

As discussed by previous studies \citep[e.g.][]{murphy:2014}, the time-delay measurements $\tau(t)$ are related in a simple way to the stellar radial velocity $v_{\rm rad}$
$$
\tau(t) = -\dfrac{1}{c}\int_0^t v_{\rm rad}(t') \mathrm{d}t',
$$
where $c$ is the speed of light. Applying this equation to the orbital radial velocity curve obtained in Sect.~\ref{sec:binary} (with the corrected semi-amplitude found in Sect.~\ref{sec:zdi}) yields the predicted time delay variation shown by the dashed line in Fig.~\ref{fig:delay} (lower panel). Both the phase and shape of this curve are in good agreement with the experimental time-delay measurements, confirming that the light travel time effect is the most likely explanation of the long-term cyclic changes of the photometric rotational variability of \dra. We also attempted to derive the orbital parameters $K$, $e$, $\omega$ directly from the photometric time-delay measurements. This approach yielded $K=42.1\pm6.7$~\kms, $e=0.674\pm0.079$, and $\omega=133\fdg0\pm3\fdg4$. The resulting orbital semi-amplitude is significantly higher than the one inferred from hydrogen lines in Sect.~\ref{sec:binary}, confirming that the latter radial velocity indicator is biased. On the other hand, the photometric radial velocity semi-amplitude is consistent within 1.2$\sigma$ with the estimate obtained from modelling Fe lines in Sect.~\ref{sec:zdi}, but is considerably less precise compared to the latter.

\subsection{Multi-line profile analysis}
\label{sec:lsd}

\subsubsection{Model atmosphere and mean abundances}

Two previous spectroscopic studies derived atmospheric parameters and analysed chemical composition of \dra. \citet{kuschnig:1998} adopted $T_{\rm eff}=12500$~K, $\log g=4.0$ and obtained surface abundance distributions of He, Si, Ti, Cr, and Fe with the Doppler imaging method. This combination of atmospheric parameters and chemical abundance maps allowed \citet{prvak:2015} to reproduce photometric variability of \dra\ from UV to near-infrared. \citet{sikora:2019} determined $T_{\rm eff}=11630$~K and $\log g=4.11$ from spectroscopic and photometric observations but estimated $\log g=3.85$ from the comparison with stellar evolutionary models. They also provided abundances of Si, Ti, Cr, and Fe. This abundance analysis was based on the same set of Narval observations as analysed in our study, but appears to have been carried out using a single high-quality observation since the authors do not mention correcting for the orbital radial velocity variation which would have been necessary to obtain a phase-averaged spectrum.

Taking into account results of the previous studies, we adopted $T_{\rm eff}=12000$~K, $\log g=4.0$ and calculated a stellar model atmosphere with these parameters using the {\sc LLmodels} code \citep{shulyak:2014}. These atmospheric parameters together with the abundances reported by \citet{sikora:2019} were employed to obtain a preliminary line list from the VALD3 data base \citep{ryabchikova:2015}. We constructed a mean spectrum of \dra\ by removing the orbital radial velocity offsets predicted by our spectroscopic orbital solution and averaging all 22 observations. The resulting mean spectrum was compared with the theoretical synthetic spectrum calculated with the {\sc Synth3} code \citep{kochukhov:2007d} for the parameters listed above and $v_{\rm e}\sin i=81.9$~\kms\ \citep{sikora:2019}. Abundances of several elements with a large number of absorption features were then adjusted to reduce the discrepancy between the average observed spectrum and calculations. This analysis yielded the following set of mean abundances: $\log (N_{\rm He}/N_{\rm tot})=-2.40$, $\log (N_{\rm Si}/N_{\rm tot})=-3.52$, $\log (N_{\rm Ti}/N_{\rm tot})=-6.90$, $\log (N_{\rm Cr}/N_{\rm tot})=-5.58$, and $\log (N_{\rm Fe}/N_{\rm tot})=-4.09$ with an uncertainty of about 0.2~dex. A new atomic line list was extracted from VALD3 using this revised abundance table.

\subsubsection{Least-squares deconvolved profiles}

\begin{figure}
\centering
\figps{\hsize}{0}{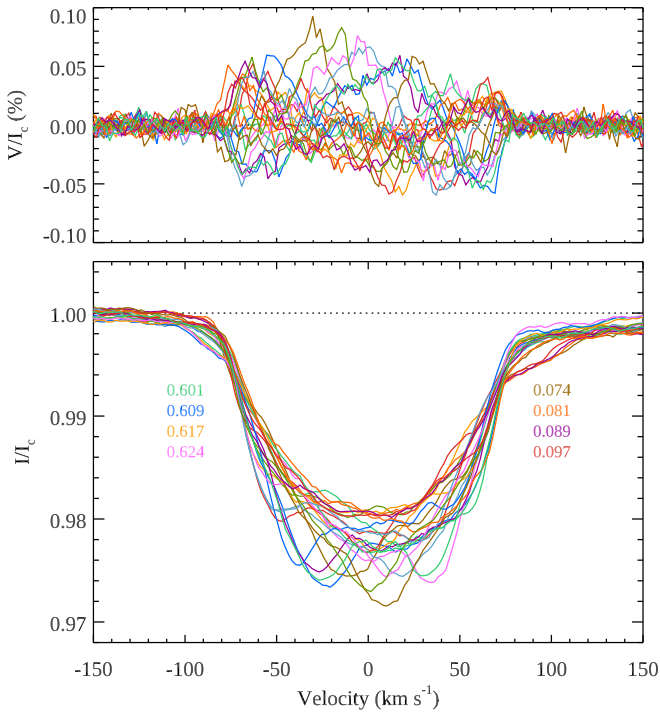}
\caption{Stokes $V$ (upper panel) and $I$ (lower panel) LSD profiles derived from all metal lines. Spectra are shown in the reference frame of the primary. Orbital phases of the LSD profiles showing the largest blue- or red-wing distortions are indicated in the lower panel.}
\label{fig:lsd}
\end{figure}

Due to a large rotational line broadening and moderate magnetic field strength, circular polarisation spectra of \dra\ do not readily show Zeeman signatures except in the few strongest lines. In this situation it is convenient to apply the least-squares deconvolution (LSD, \citealt{donati:1997}; \citealt*{kochukhov:2010a}) procedure, which efficiently co-adds intensity and polarisation spectra of many individual lines into high-quality mean line profiles. In this work we used the {\sc iLSD} code \citep{kochukhov:2010a} and the VALD3 line list tailored for \dra\ for the LSD profile calculation. Initially, we considered all metal lines deeper than 5 per cent of the continuum, excluding wavelength regions contaminated by telluric absorption and affected by broad wings of the hydrogen Balmer lines. In practice, the wavelength range was restricted to 4000--7850~\AA. These selection criteria yielded 1260 metal lines and resulted in Stokes $V$ LSD profiles with a typical uncertainty of $6.3\times10^{-5}$ per 2~\kms\ velocity bin. This set of Stokes $V$ spectra and the corresponding Stokes $I$ profiles are illustrated in Fig.~\ref{fig:lsd}. The LSD line mask normalisation adopted for this calculation (see \citealt{kochukhov:2010a} for details) was $\lambda_0=5000$~\AA\ and $z_0=1.2$. 

As evident from Fig.~\ref{fig:lsd}, variable Stokes $V$ signal with an amplitude of about $\pm0.05$ per cent of the unpolarised continuum is detected well above the noise level in all observations. No signatures are seen in the LSD profiles computed from the diagnostic null spectra. The Stokes $I$ profiles exhibit a strong rotational modulation, primarily due to spots of Fe and Cr which dominate (941 out of 1260 lines) the metal line mask. In addition to this prominent variability, we detected weak variable absorption outside the $\pm v_{\rm e}\sin i$ velocity range, which correlated with the orbital rather than rotational phase. For example, the group of four profiles with a conspicuous red-wing absorption component corresponds to the observations obtained in the 0.074--0.097 orbital phase interval when the primary (component Aa) had a negative velocity with respect to the centre of mass and the secondary (component Ab) had a positive velocity offset (see Fig.~\ref{fig:orbit}). Conversely, another group of four observations, obtained in the 0.601--0.624 orbital phase interval, shows depression on the blue side of the line when the secondary was predicted to have a negative velocity offset relative to the primary. Therefore, distortions seen in the outer wings of the Stokes $I$ profiles are consistent with contribution of the secondary star. This indicates that \dra\ is an SB2 system. Unfortunately, our observations have a poor coverage of the 127.9-day binary orbit, precluding us from pursuing further binary-star analysis such as deriving radial velocities of the secondary and spectral disentangling.

The LSD profiles calculated with the complete metal line mask allow one to achieve the highest precision in detecting Zeeman signatures. However, such profiles are unsuitable for detailed quantitative modelling of magnetic field and starspot geometries because the line strength modulation influencing both $I$ and $V$ profiles is a result of compound effect of several chemical elements with different surface distributions. The modelling task can be accomplished only by considering LSD profiles of individual chemical elements. To this end, we derived three additional sets of LSD profiles for Fe, Si, and Cr. The corresponding LSD masks included 828, 103, and 113 lines, respectively, with the remaining metal lines treated as a background with the multi-profile LSD approach \citep{kochukhov:2010a}. The Fe and Si profiles obtained using this method are modelled in Sect.~\ref{sec:zdi} to derive both magnetic field topology and abundance distributions of these elements. Chromium lines, being less numerous than those of Fe and, on average, weaker than those of Si, yield rather noisy Stokes $V$ LSD profiles. For this reason, we use only the Stokes $I$ profiles of this element for abundance spot mapping.

\begin{figure}
\centering
\figps{\hsize}{0}{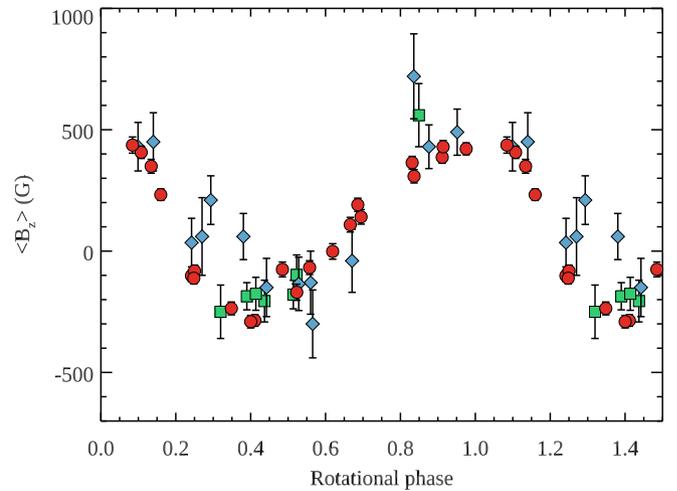}
\caption{Mean longitudinal magnetic field of \dra\ as a function of rotational phase. Measurements derived in this study (circles) are compared with the MuSiCoS measurements reported by \citet[][squares]{sikora:2019a} and Balmer-line measurements by \citet[][diamonds]{landstreet:1977}.}
\label{fig:bz}
\end{figure}

\subsubsection{Longitudinal magnetic field}

\begin{figure*}
\centering
\figps{\hsize}{0}{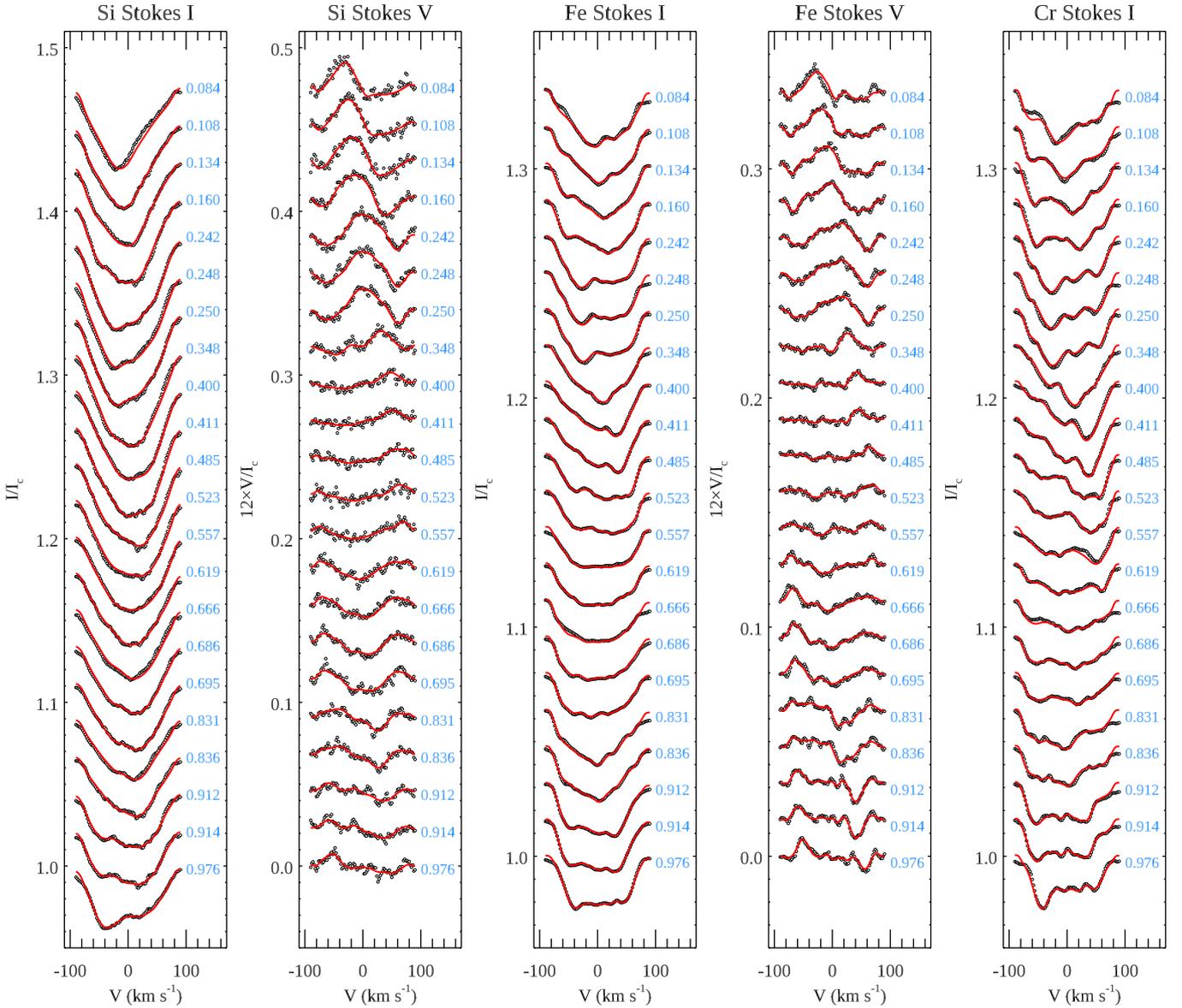}
\caption{Comparison of the observed (symbols) and model (solid lines) Stokes $IV$ Si, Fe, and Cr LSD profiles. The spectra are offset vertically, with the rotational phase indicated next to each profile. The Stokes $V$ spectra are magnified by a factor 12 relative to Stokes $I$.}
\label{fig:zdi_fit}
\end{figure*}

The mean longitudinal magnetic field, \bz, was determined from the first moment of the Stokes $V$ LSD profile normalised by the equivalent width of Stokes $I$ \citep{wade:2000,kochukhov:2010a}. Numerical integration required for the evaluation of line profile moments was carried out in the $\pm$\,90~\kms\ velocity interval after removing the orbital radial velocity variation. The resulting \bz\ measurements obtained from the LSD profiles derived using all metal lines are provided in Table~\ref{tab:obs}. Figure~\ref{fig:bz} illustrates these measurements as a function of rotational phase together with \bz\ measurements reported in earlier studies. According to more precise Narval measurements, \dra\ exhibits a smooth, approximately sinusoidal variability of the mean longitudinal magnetic field between about $-290$ and +440~G. Our measurements agree within uncertainties with the results derived by \citet{sikora:2019a} from the same data, so we do not pursue further quantitative analysis of the \bz\ phase curve of \dra. At the same time, it may be noted that all investigations, including ours, are slightly underestimating the field strength of \dra\ because the spectral contribution of the secondary is not removed from Stokes $I$ prior to \bz\ calculation.

\subsection{Zeeman Doppler imaging}
\label{sec:zdi}

\begin{figure*}
\centering
\figps{\hsize}{270}{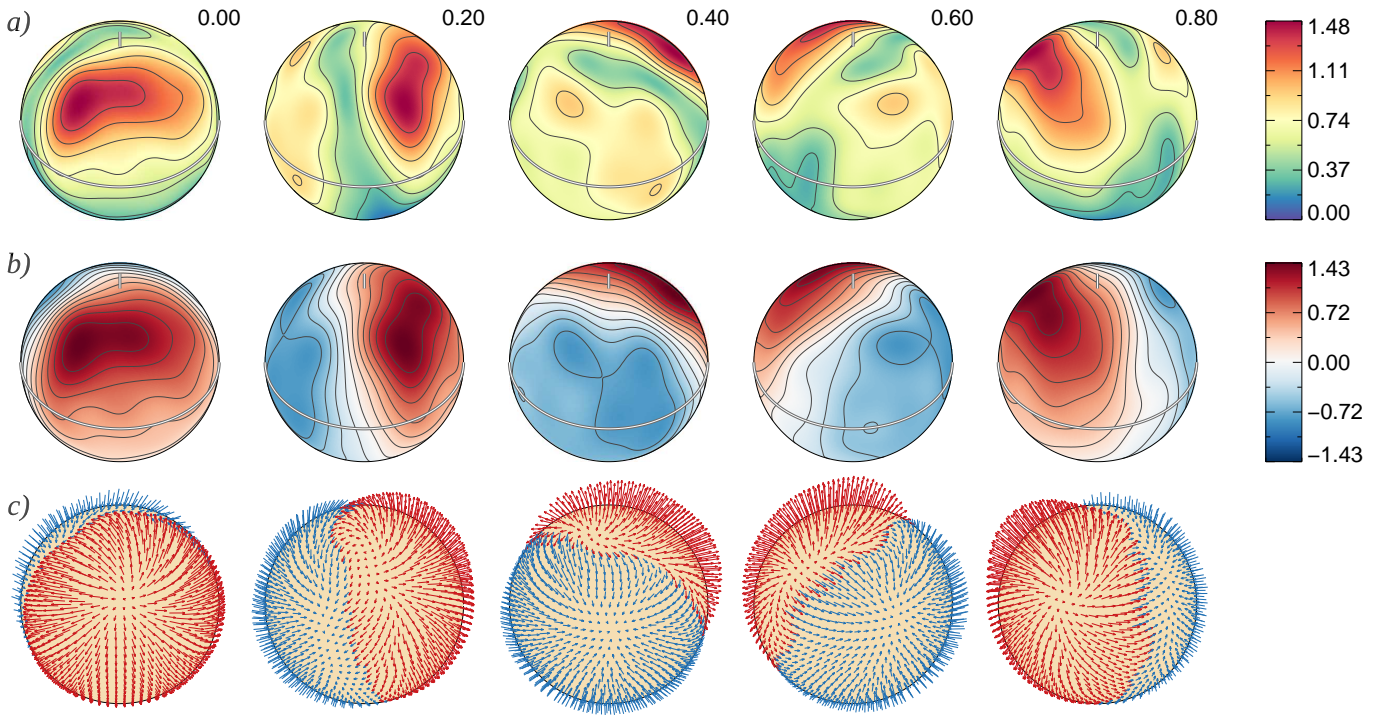}
\caption{Magnetic field topology of \dra\ derived from the Si LSD profiles. The spherical plots show maps of the total magnetic field strength (a), the radial field component (b), and the field vector orientation (c) at five rotational phases, which are indicated next to each column. The contours over spherical maps in panels (a) and (b) are plotted with a step of 0.2 kG. The stellar equator and rotational pole are indicated with double lines. The side colour bars give the field strength in kG. }
\label{fig:zdi_si}
\end{figure*}

\begin{figure*}
\centering
\figps{\hsize}{270}{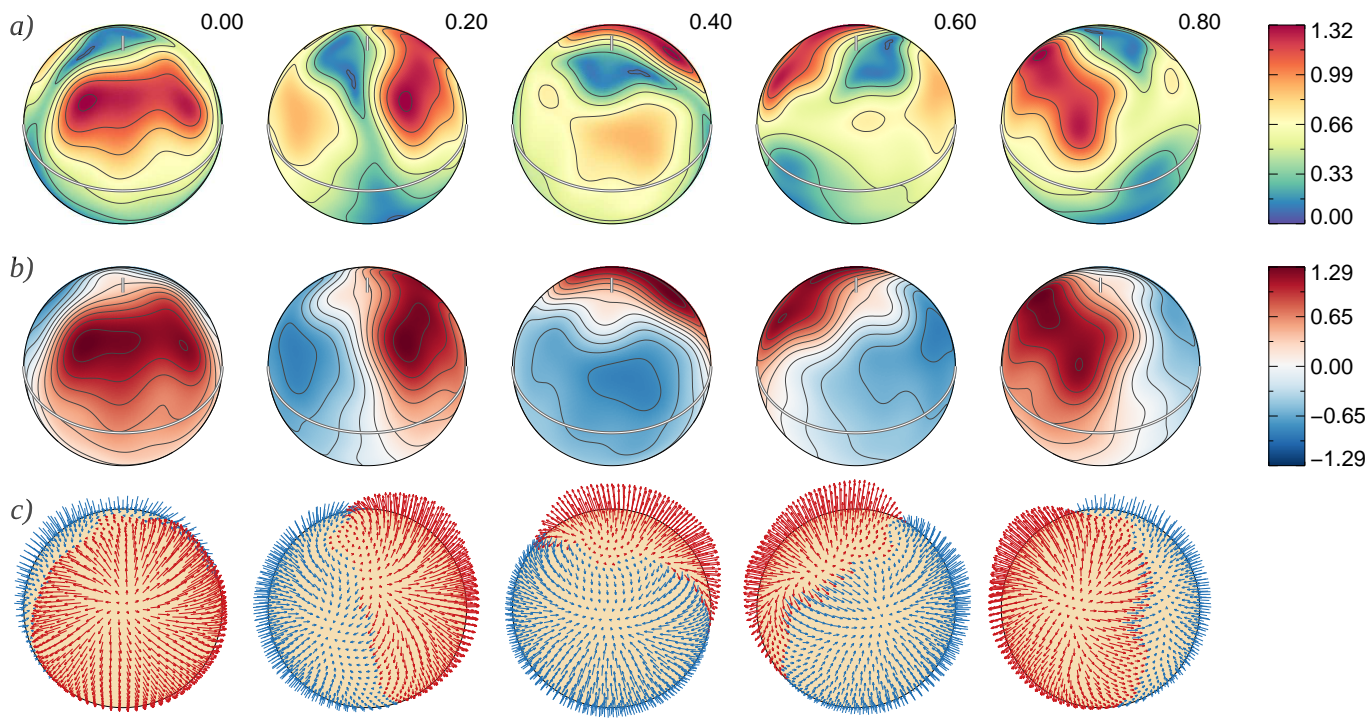}
\caption{Same as Fig.~\ref{fig:zdi_si} for the magnetic field topology derived from the Fe LSD profiles.}
\label{fig:zdi_fe}
\end{figure*}

\begin{figure*}
\centering
\figps{15.5cm}{0}{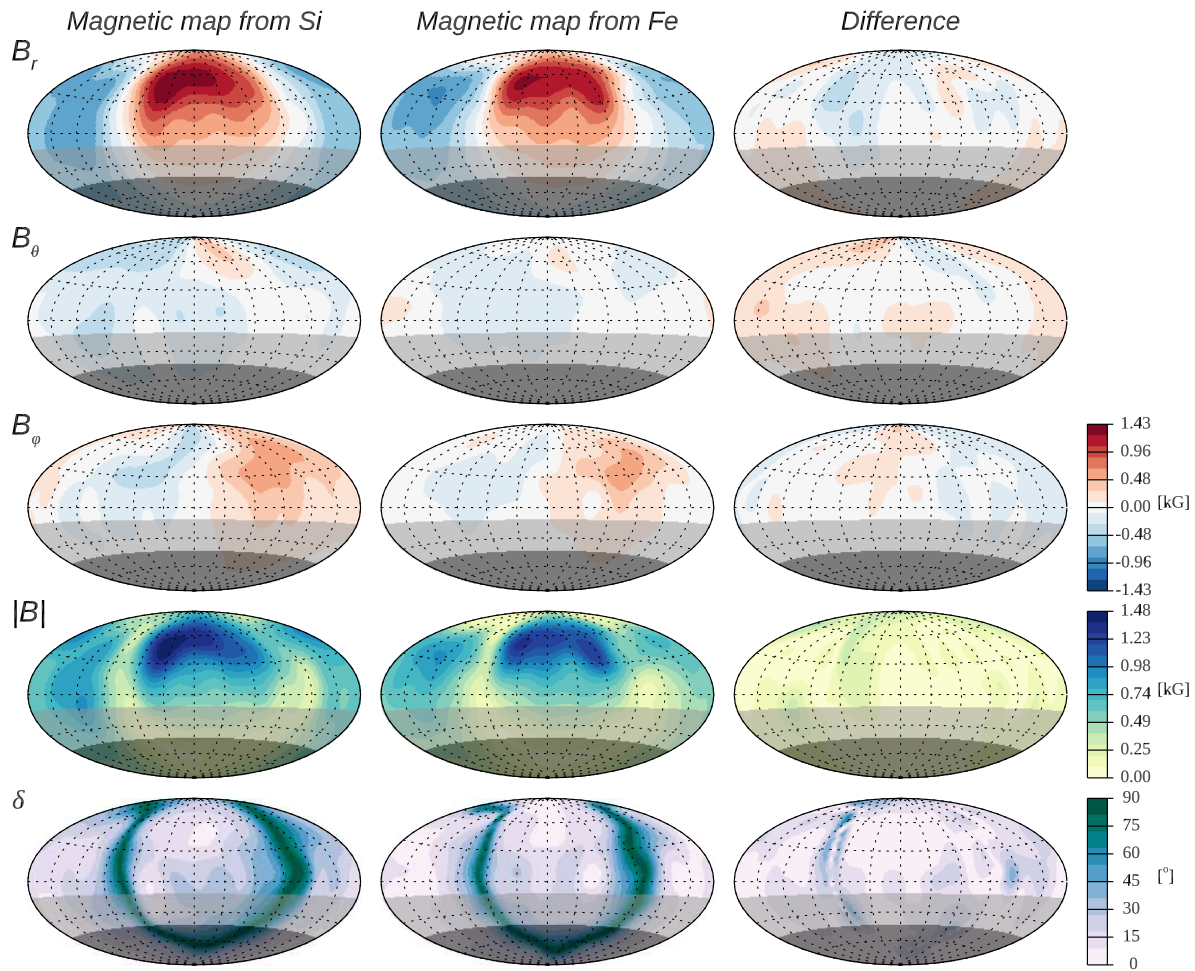}
\caption{Comparison of the magnetic field maps obtained from the Si and Fe LSD profiles. The rows show (from top to bottom) maps of the radial, meridional, azimuthal field components, the field modulus, and the field inclination relative to the surface normal. 
All maps are shown in the Hammer-Aitoff projections with the central meridian corresponding to phase zero. The shaded bottom part of the maps indicates latitudes that are invisible (dark grey) or have a relative visibility of less than 25 per cent (light grey) for the adopted inclination angle $i=48\degr$. 
}
\label{fig:dif}
\end{figure*}

\begin{figure}
\centering
\figps{7.5cm}{0}{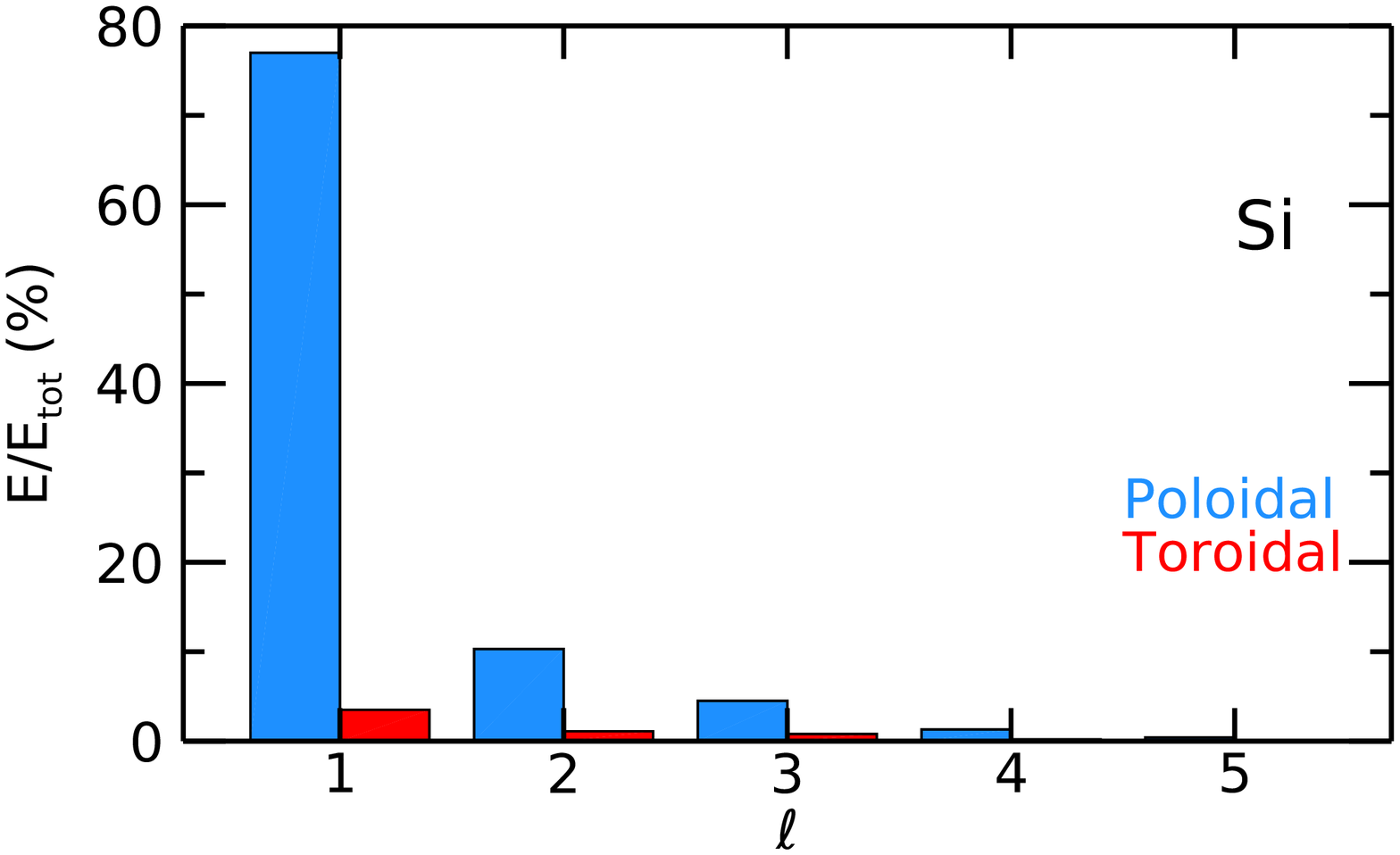}\vspace*{5mm}
\figps{7.5cm}{0}{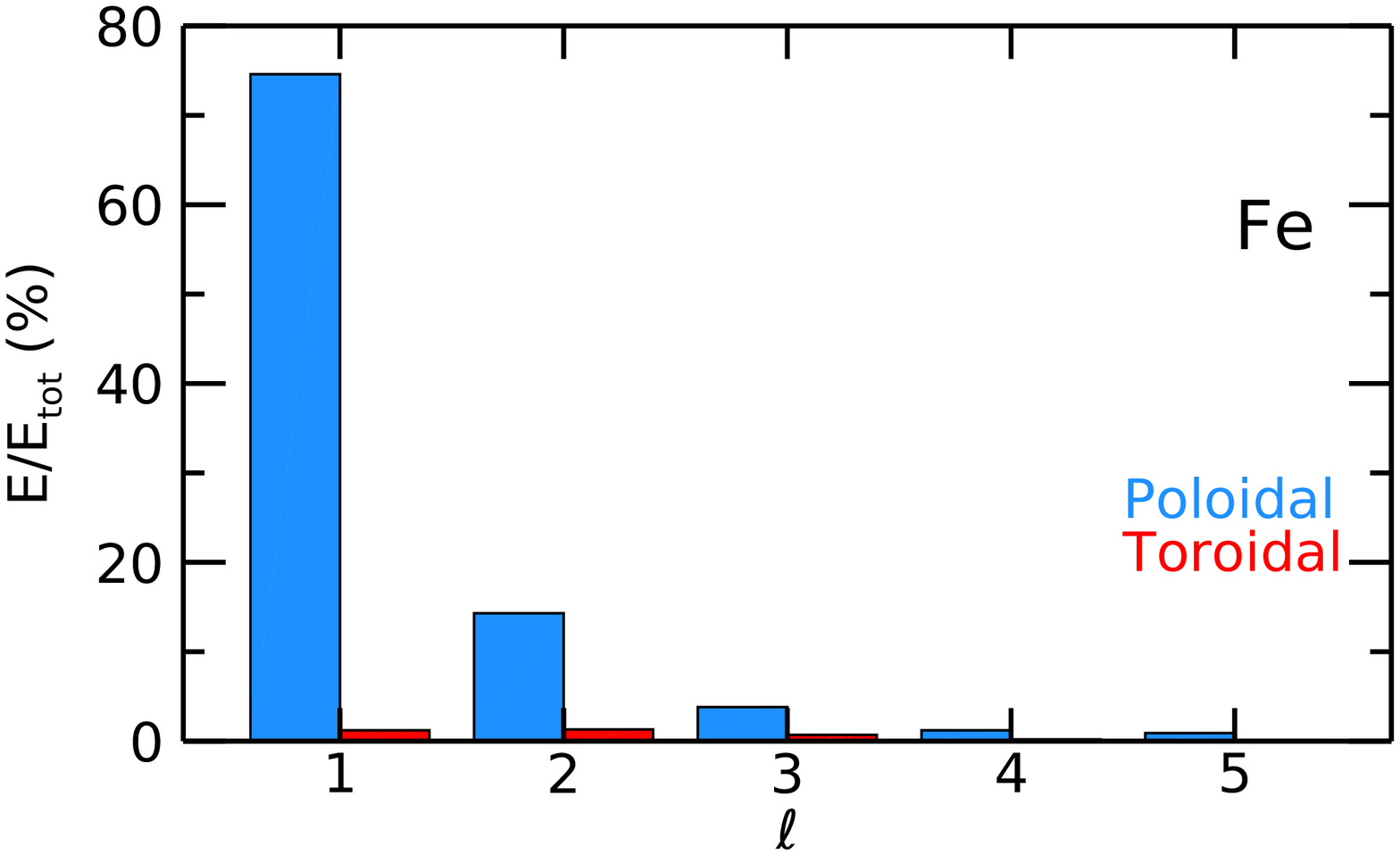}
\caption{Relative energies of the poloidal and toroidal harmonic components as a function of the angular degree $\ell$ for the magnetic field topologies reconstructed from the Si (top) and Fe (bottom) LSD profiles.}
\label{fig:harm}
\end{figure}

The magnetic field geometry and chemical abundance maps of \dra\ were reconstructed with the help of the Zeeman Doppler imaging (ZDI) code {\sc InversLSD} introduced by \citet{kochukhov:2014} and subsequently used by \citet{kochukhov:2017a,kochukhov:2019}, \citet*{rosen:2015}, and \citet{oksala:2018}. This tomographic mapping procedure allows one to interpret rotational modulation of the LSD Stokes parameter profiles of magnetic stars based on detailed polarised radiative transfer calculations in a realistic model atmosphere. This is accomplished by calculating a grid of local Stokes spectra, including all relevant absorption lines for the entire observed wavelength interval, and a range of magnetic field strengths, field vector inclinations, chemical abundances, and limb angles and then applying the LSD procedure to these calculations using the same line mask and relative weights as in the treatment of observations. The resulting grid of theoretical local Stokes parameter profiles is tabulated once and then used by the inversion module, which interpolates within this local profile table during calculation of the disk-integrated spectra and their derivatives. This approach allows us to separate time-consuming detailed polarised spectrum synthesis, usually carried out at a supercomputing facility, from the actual ZDI inversions for different sets of nuisance parameters, which require relatively modest computational resources.

In this study we reconstructed the magnetic field geometry from the Stokes $I$ and $V$ profiles of Fe and Si. Chemical abundance maps were derived simultaneously for both elements. In addition, we obtained a non-uniform distribution of Cr from its Stokes $I$ LSD profiles. For each of the three chemical elements relevant abundance ranges were covered with a step of 0.25~dex during calculation of local LSD profiles. A step of 200 G was employed for the magnetic field strength while both the angle between the field vector and the line of sight and limb angle parameter spaces were sampled with 15 values each. The same {\sc LLmodels} atmosphere with $T_{\rm eff}=12000$~K and $\log g=4.0$, the mean abundances discussed above, and a complete VALD3 line list comprising about 5000 transitions in the 4000--7850~\AA\ wavelength interval were adopted for the spectrum synthesis calculations.

The global magnetic field geometry of \dra\ was parametrised using a spherical harmonic expansion \citep{donati:2006b,kochukhov:2014}, which included both poloidal (potential) and toroidal (non-potential) terms. Three independent sets of the spherical harmonic coefficients, $\alpha_{\ell m}$, $\beta_{\ell m}$, $\gamma_{\ell m}$, were employed to describe the radial poloidal, horizontal poloidal, and horizontal toroidal field components respectively. The expansion was truncated at the maximum angular degree $\ell_{\rm max}=10$, which is sufficient to describe global magnetic fields of Ap stars. A harmonic penalty function described by \citet{morin:2008} and \citet{kochukhov:2014} was applied to prevent the inversion procedure from introducing unnecessarily complex magnetic field distributions. Chemical abundance maps were prescribed using a 1876-element surface grid and regularised with the Tikhonov method \citep{piskunov:2002a}. Both the abundance and magnetic field regularisation parameters were adjusted following the procedure outlined by \citet{kochukhov:2017} to achieve a balance between reaching a satisfactory fit to observations and avoiding spurious small-scale surface features. 

An inclination angle of $i=48\degr$ was adopted for all DI and ZDI inversions in this study. This parameter was derived by \citet{sikora:2019a} from the stellar rotational period, radius, and \vsini. The uncertainty of this inclination angle estimate is about $\pm5\degr$.

\subsubsection{Refinement of the stellar and orbital parameters}

As discussed above, the blending of hydrogen lines of the Ap primary by the secondary star in the inner spectroscopic binary introduces a systematic bias in the radial velocity measurements, yielding an underestimated orbital velocity amplitude. Metal lines are intrinsically narrower, allowing one to better isolate spectral signature of the primary and potentially obtain a more accurate orbital amplitude. However, metal lines exhibit strong rotational modulation due to chemical spots on the primary. Thus, in practice, a refinement of the orbital parameters using metal lines is only possible if one can reproduce Stokes $I$ profile shapes in every observation. This necessitates modelling of abundance spots with a DI code. To this end, we carried out chemical spot inversions using the Fe Stokes $I$ LSD profiles for a two-dimensional parameter grid spanning a range of $K$ and $\gamma$ values while keeping all other orbital elements fixed. A well-defined $\chi^2$ minimum was found at $K=33.87\substack{+0.25 \\ -0.17}$~\kms\ and $\gamma=-16.44\substack{+0.10 \\ -0.08}$~\kms. This centre-of-mass velocity differs little from the orbital solution based on the hydrogen-line radial velocity measurements. On the other hand, the orbital semi-amplitude found using Fe lines is 26 per cent higher than the one inferred from H lines. This orbital parameter derived from Fe lines is more accurate than the orbital amplitude obtained from H lines since spectroscopic contribution of the primary can be better isolated using intrinsically narrower metal lines. Consequently, we adopt the orbital parameters determined in this section for all subsequent ZDI calculations.

Aiming to improve the \vsini\ estimate, we carried out ZDI inversions for a range of projected rotational velocities and compared the resulting fit quality. This analysis was performed for both Si and Fe LSD profiles, yielding \vsini\,=\,$82.0\pm0.5$~\kms\ and \vsini\,=\,$81.0\pm0.5$~\kms\ for these two elements respectively. We adopted \vsini\,=\,$81.5\pm0.5$~\kms\ as a compromise value. This determination of the projected rotational velocity of \dra\ is consistent with \vsini\,=\,$81.9\pm1.6$~\kms\ found by \citet{sikora:2019} from the conventional spectrum synthesis assuming uniform chemical abundance distributions.

\subsubsection{Magnetic field topology}

Results of the tomographic reconstruction of the magnetic field structure of \dra\ using the Si and Fe LSD profiles are illustrated in Figs.~\ref{fig:zdi_fit}--\ref{fig:harm}. The first of these figures compares the observed and model Stokes profiles. Figs.~\ref{fig:zdi_si} and \ref{fig:zdi_fe} present spherical maps of the field strength, the radial field component, and the field vector orientation obtained from modelling Si and Fe lines, respectively. In both cases we derived a distorted dipolar topology characterised by a large obliquity relative to the stellar rotational axis and a significant asymmetry between the positive and negative magnetic poles. The former is represented by an arc-like structure in which the field strength reaches 1.32--1.48~kG. The latter is much more diffuse, with a field strength below 1~kG.

Fig.~\ref{fig:dif} provides a closer look at the inter-agreement of the magnetic field maps derived from the mean profiles of the two chemical elements. This figure shows that the results of independent magnetic inversions are broadly consistent with each other despite a noticeably different pattern of the Stokes $I$ and $V$ profile variability (see Fig.~\ref{fig:zdi_fit}). Calculating the mean absolute values of the difference maps in the northern stellar hemisphere one can infer that the radial, meridional, and azimuthal field components of the two maps are consistent to within 113, 112, and 104~G, respectively. This corresponds to 7.4--8.1 per cent of the peak surface field strength. Likewise, the field modulus distributions agree to within 118~G (8.4 per cent of the peak field) and the local field inclination is discrepant by 11.6\degr\ on average. These figures can be considered as an estimate of the internal precision of the magnetic inversions applied in this study.

The harmonic content of the reconstructed magnetic field maps is assessed in Fig.~\ref{fig:harm}. We find that most (75--77 per cent) of the magnetic field energy is concentrated in the dipolar component and that the field structure is predominately poloidal (all toroidal spherical harmonic modes combined contribute only 3.5--5.6 per cent of the total field energy) and does not contain a significant small-scale field (all modes with $\ell>5$ combined contribute less than 2 per cent of the field energy).

\begin{figure*}
\centering
\figps{\hsize}{270}{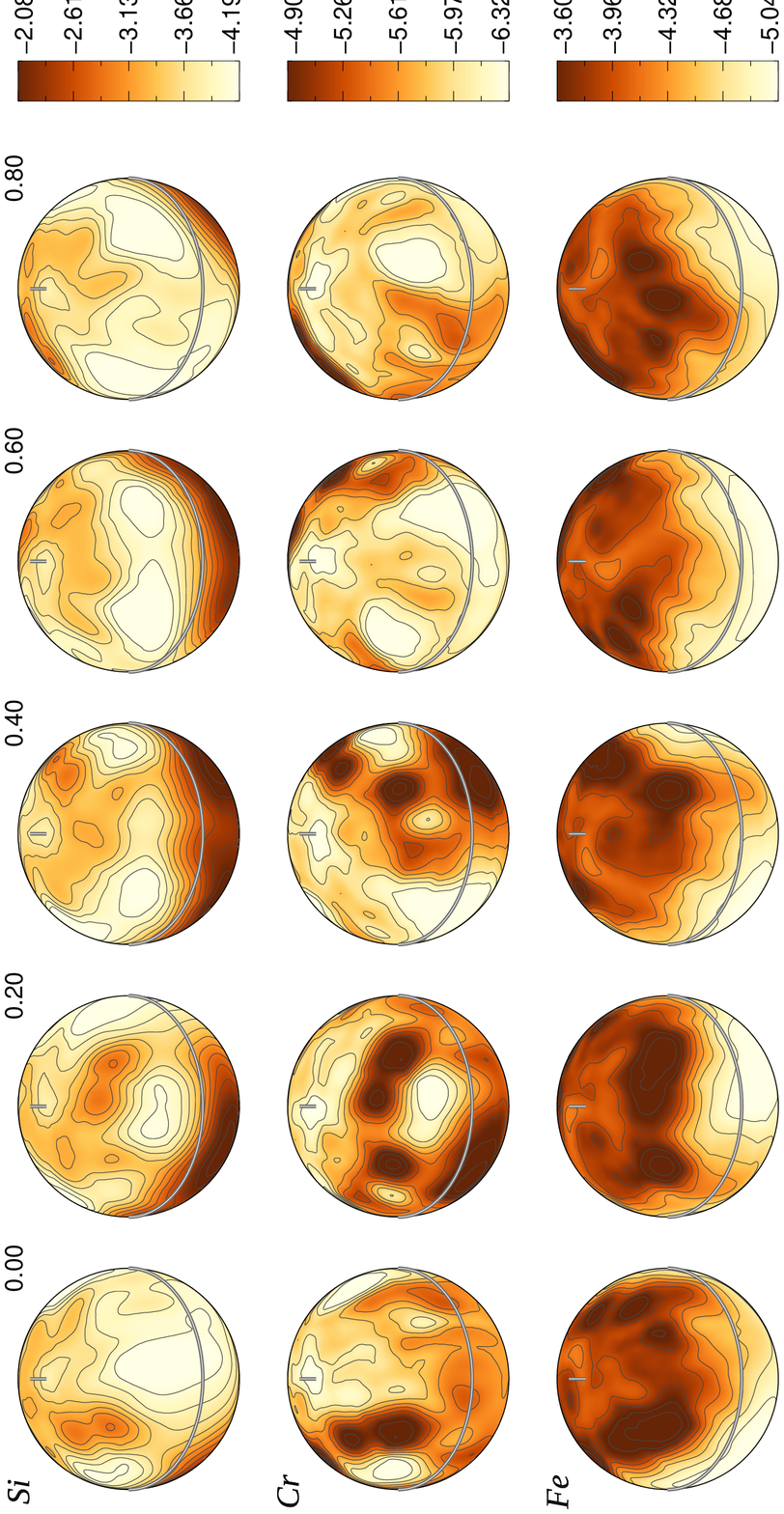}
\caption{Surface distributions of Si, Fe, and Cr derived from the LSD profiles of these elements. The fractional element abundances are indicated by the side bars in the $\log (N_{\rm el}/N_{\rm tot})$ units, with the minimum and maximum values corresponding to 5 and 95 percentiles, respectively. The contours over spherical maps are plotted every 0.2 dex.}
\label{fig:zdi_abn}
\end{figure*}

To conclude the ZDI analysis, we assessed the feasibility of reproducing spectropolarimetric observations of \dra\ with a canonical oblique dipolar topology \citep[e.g.][]{sikora:2019a} or with a superposition of dipolar and general non-axisymmetric quadrupolar field geometries \citep{bagnulo:2002}. In these tests, conducted using the Fe Stokes $I$ and $V$ LSD profiles, the magnetic field geometry was reconstructed setting $\ell_{\rm max}=1$ or 2 and adopting $\beta_{\ell m}=\alpha_{\ell m}$, $\gamma_{\ell m}=0$ spherical harmonic expansion (see \citealt{kochukhov:2014} and \citealt{kochukhov:2016a} for detailed discussions of different harmonic parameterisations). No magnetic field regularisation was applied and the Fe abundance distribution was reconstructed simultaneously with the magnetic field map. With these restrictions {\sc InversLSD} was not able to obtain a satisfactory fit to the Fe Stokes $V$ profiles, with the final standard deviation remaining 55--68 per cent higher than for the inversion based on the general high-$\ell$ harmonic parameterisation normally adopted in ZDI. This confirms that the global distortion of the dipolar field as well as higher-order harmonic perturbations are essential for reproducing observations of \dra\ and likely reflect real deviations of the surface field structure from a pure dipole.

\subsubsection{Chemical spot distributions}

Surface abundance distributions of Si and Fe were reconstructed simultaneously with the magnetic field maps from the Stokes $I$ and $V$ profiles of these elements. The resulting chemical spot maps are presented in Fig.~\ref{fig:zdi_abn}. This figure also shows the surface distribution of Cr recovered from the mean intensity profiles of this element. The corresponding observed and computed Stokes $I$ spectra are shown in the right-most column of Fig.~\ref{fig:zdi_fit}. This DI reconstruction of the Cr spot geometry was carried out using the magnetic field map obtained by averaging the field distributions inferred in the ZDI analyses of Si and Fe.

Significant surface inhomogeneities are found for all three chemical elements. Si shows the smallest ($\sim$\,1~dex) variation across the northern stellar hemisphere. Fe and Cr exhibit stronger inhomogeneities, with the abundance contrasts exceeding 1.5~dex. The two studied Fe-peak elements possess similar surface distributions over a significant fraction of the stellar surface, but Cr shows larger and higher contrast overabundance spots compared to Fe. At the same time, Fe also exhibits a more complex surface distribution with an additional series of spots (phases 0.6--0.8) not visible in the Cr map.

For all three chemical elements inversions yield major under- or overabundance areas below the stellar rotational equator. This part of the stellar surface is poorly sampled by observations owing to a moderate inclination of \dra. In this situation, abundance mapping in the sub-equatorial  regions is more prone to artefacts, including those possibly related to unaccounted blending by the secondary component. Using the projected surface area summed over all rotational phases as a relative visibility proxy, one can conclude that the latitudes below $-11.8$\degr\ contribute less than 25 per cent to the disk-integrated profiles compared to the high-latitude zones which have the best visibility. Therefore, the prominent stellar abundance features reconstructed in this study below the rotational equator are less reliable than the structures at higher latitudes.

\section{Conclusions and discussion}
\label{sec:disc}

This study presented a comprehensive investigation of the rotational variability, binarity, magnetism, and surface structure of the bright Ap star \dra. We significantly improved precision of the stellar rotational period, finding $P_{\rm rot}=1.71650213(21)$ d, based on the photometric observations collected for this star by the \textit{TESS} satellite over a period of nearly one year. An in-depth analysis of the rotational photometric phase curve revealed its modulation with the 127.9-d period corresponding to the motion of the Ap star in a spectroscopic binary with a lower mass, fainter companion. We successfully interpreted this long-term light curve modulation as the light travel time effect \citep{shibahashi:2012,murphy:2014} -- the first time that this phenomenon has been observed to affect the stellar rotational variability.

Using a set of high-resolution spectra we obtained new radial velocity measurements for \dra\ and revised parameters of its spectroscopic binary orbit taking into account all available high-precision radial velocity determinations. We also reported the first detection of spectroscopic contribution of the secondary in metal line profiles, thereby changing the status of \dra\ from an SB1 to SB2 system.

\begin{figure*}
\centering
\figps{15.5cm}{0}{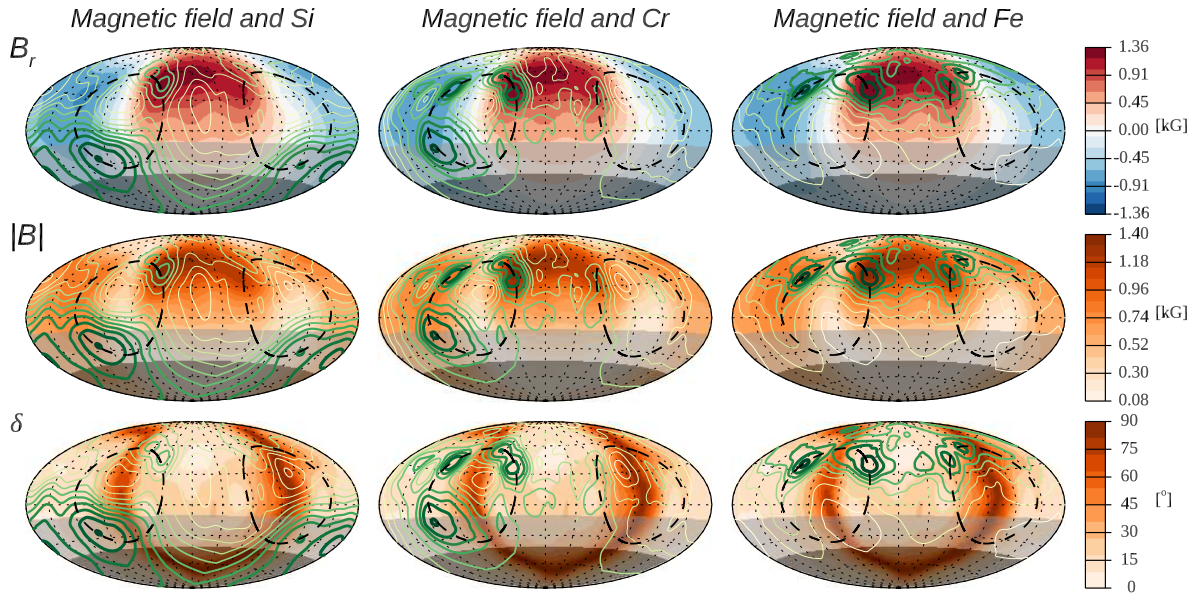}
\caption{Comparison of the magnetic field topology of \dra\ with the surface abundance distributions of Si, Cr, and Fe. The radial field component, the field modulus, and the local field inclination for the average magnetic field map are illustrated from top to bottom, similar to Fig.~\ref{fig:dif}. Chemical abundance spots are shown with contours plotted every 0.25~dex. The darker and thicker contours correspond to zones with a higher element abundance. Dashed lines trace two small circles on which overabundance spots are found.}
\label{fig:fld_abn}
\end{figure*}

We derived mean intensity and polarisation profiles of Fe, Cr, and Si with the help of the LSD technique. This multi-line analysis is a necessary prerequisite to studying magnetic field topology of \dra\ since its relatively weak magnetic field and rapid rotation does not allow one to use individual lines for magnetic diagnostic. We were able to derive mean circular polarisation signatures with a quality sufficient for detailed modelling from the lines of Fe and Si. A magnetic tomographic mapping technique was then applied to these observations in order to obtain vector magnetic field maps simultaneously with chemical spot distributions. This analysis yielded two independent magnetic field distributions, which agreed at the level of $\sim$\,10 per cent of the peak field strength. The surface field topology of \dra\ was found to be primarily poloidal and dipolar, with small but statistically significant contributions of higher-order poloidal harmonic components. This magnetic field topology is characterised by a maximum local field modulus of 1.39~kG and a surface-averaged field strength of 0.54~kG, making it one of the lowest-field Ap stars studied with ZDI. In general, the surface magnetic field properties of \dra\ are unremarkable compared to other Ap stars studied with magnetic mapping (see \citealt{kochukhov:2019} and references therein).

We reconstructed chemical spot distributions of Si and Fe simultaneously with mapping magnetic field using the Stokes $I$ and $V$ LSD profiles of these elements. In addition, distribution of Cr was recovered from the Stokes $I$ LSD profile of this element keeping the magnetic field map fixed. This set of three chemical maps is small compared to multi-element distributions available for some Ap stars (e.g. \citealt{kochukhov:2004e,luftinger:2010,nesvacil:2012}; \citealt*{silvester:2014a}). However, an advantage of the present study is that a detailed map of magnetic field is recovered self-consistently from the same observational material and that this magnetic field is relatively weak, simplifying both reconstruction and theoretical interpretation of the chemical abundance maps. In addition, \dra\ is one of the fastest rotating Ap stars to which modern ZDI analysis was applied, yielding higher resolution surface maps compared to DI results typically found in the literature (see summary in \citealt{kochukhov:2017}).

Spherical maps of \dra\ (Figs.~\ref{fig:zdi_si}, \ref{fig:zdi_fe}, \ref{fig:zdi_abn}), which are often analysed in the discussion of ZDI and DI results for Ap stars, do not readily reveal a relation between the magnetic field topology and chemical abundance distributions. To investigate this relation in more detail, we plotted abundance maps over the magnetic field geometry using the Hammer-Aitoff map projection (Fig.~\ref{fig:fld_abn}). It is evident that spot maps are morphologically more complex than the magnetic field structure and do not exhibit a simple correlation with the local field inclination as predicted by many theoretical atomic diffusion models \citep[e.g.][]{leblanc:2009,alecian:2010,alecian:2015,kochukhov:2018a}. Nevertheless, one can note that one of the abundance spots in all three studied chemical maps coincides with the same strong-field region where the field vector is directed outwards.

Despite the apparent lack of a simple abundance-magnetic field correlation, we recognised an intriguing regularity of the surface locations of chemical spots relative to the stellar rotational and magnetic field topologies. The spots are found along the two small circles, which have opening angles of $\approx$\,45\degr, are located $\approx$\,180\degr\ apart and centred at a latitude of $\approx$\,$+10$\degr\ at the magnetic equator. These small circles, highlighted by the dashed lines in Fig.~\ref{fig:fld_abn}, appear to account for all main overabundance features of the three studied elements, including spots below the rotational equator which we previously deemed to be less reliable. There is a tendency for the overabundance spots to be found along these circles in one of the four quadrants defined by the intersection of the rotational and magnetic equators. At the same time, spots typically avoid the magnetic equator (except for one high-latitude feature in the Fe map) and populate these quadrants differently depending on the element. Nevertheless, a systematic behaviour of spot locations is clearly present, hinting at the importance of stellar rotation alongside the magnetic field for the atomic diffusion. Further analysis of the DI results obtained for other Ap stars with the graphical representation similar to Fig.~\ref{fig:fld_abn} is necessary to assess if the star spot behaviour uncovered in this study of \dra\ is a unique property of this particular star or can be found in other objects.

\section*{Acknowledgements}
OK acknowledges support by the Swedish Research Council (grant no. 2019-03548), the Royal Swedish Academy of Sciences and the Swedish National Space Agency.
This work is based on observations obtained at the \textit{Bernard Lyot} Telescope (TBL, Pic du Midi, France) of the Midi-Pyr\'en\'ees Observatory, which is operated by the Institut National des Sciences de l'Univers of the Centre National de la Recherche Scientifique of France.
The computations in this study were enabled by resources provided by the Swedish National Infrastructure for Computing (SNIC), partially funded by the Swedish Research Council through the grant agreement no. 2018-05973.
This research has made extensive use of the SIMBAD database, operated at CDS, Strasbourg, France.

\section*{Data availability}
The spectropolarimetric data underlying this article are available from the PolarBase archive (\url{http://polarbase.irap.omp.eu}).
The \textit{TESS} observations of \dra\ can be obtained from the Mikulski Archive for Space Telescopes (MAST, \url{https://mast.stsci.edu}).


\bsp
\label{lastpage}
\end{document}